Distorted *Born* diffraction tomography applied to inversing the ultrasonic field scattered by a non-circular infinite elastic tube

Philippe Lasaygues, Régine Guillermin, Jean-Pierre Lefebvre

Laboratoire de Mécanique et d'Acoustique, UPR CNRS 7051

31 chemin Joseph Aiguier,

13402 Marseille cedex 20 – France

Corresponding author:

Philippe Lasaygues

Laboratoire de Mécanique et d'Acoustique, UPR CNRS 7051

31 chemin Joseph Aiguier,

13402 Marseille cedex 20 – France

Running title: Distorted Born Diffraction Tomography

**ABSTRACT** (103 words)

This study focuses on the application of an ultrasonic diffraction tomography to non-circular 2D-cylindrical objects immersed in an infinite fluid. The *distorted Born iterative method* used to solve the inverse scattering problem belongs to the class of algebraic reconstruction algorithms. This method was developed to increase the order of application of the Born approximation (in the case of weakly contrasted media) to higher orders. This yields quantitative information about the scatterer, such as the speed of sound and the attenuation. Quantitative ultrasonic imaging techniques of this kind are of great potential value in fields such as medicine, underwater acoustics and non-destructive testing.

I.  INTRODUCTION

The results of studies on the ultrasonic characterization and imaging of elastic tubes have many potential technical applications in fields such as medicine, geophysics, underwater acoustics and non-destructive testing (NDT). However, ultrasonic wave propagation is greatly perturbed by the difference in the acoustic impedance (acoustic impedance contrast) between the scatterer and the surrounding medium (soft tissues, water or coupling gel), which results in considerable parasite events such as the refraction, attenuation and scattering of the waves. Many authors have dealt in the past with the ultrasonic characterization and imaging of elastic tubes. Their main aim has usually been to assess the thickness of the tube (as part of its geometrical conformation) and to calculate the speed of sound crossing the shell. To study the shape of elastic tubes, a sinogram (giving the diffracted angle vs. time) can be obtained from experimental or numerical data recorded in the plane perpendicular to their generator using radial or cross-section methods of measurement and imaging processes. In the present approach, this problem is viewed in terms of the scattering of ultrasonic waves by a solid cylindrical cavity. Ultimately, the aim is to be able to solve inverse scattering problems in the case of tubes, but for this to be possible, it is necessary to solve the corresponding forward problem. Several methods have been applied so far to the forward problem, including integral equation methods, hybrid FEM methods, and the geometrical theory of diffraction. The latter method provides asymptotic approximations for diffracted fields, which are valid at large distances from the diffracting body.

Actually all these approaches involve the "classical" problem of minimization of the differences between modeling data and measurements. Several strategies can be used for this purpose, which make it possible to model the forward problem and the inverse problem simply, efficiently and accurately. Based on the first-order Born approximation [1], ultrasonic tomography [2] is one of these strategies [3], which is known to be a potentially valuable method of imaging objects with a similar acoustical impedance to that of the surrounding homogeneous medium (in soft tissue characterization, [4] and underwater acoustics [5]). But difficulties arise when it is proposed to obtain quantitative tomograms using acoustical parameters such as the velocity or the attenuation of the wave, or, tomograms of more highly contrasted media (in the case of hard tissues [6 , industrial process tomography [7, 8], etc.). Finding solutions here involves either using iterative schemes [9] and/or performing extensive studies on the limitation of the first-order Born approximation [10].

In this paper, we describe the theoretical foundations of first-order Born tomography, recall the limits of this method when dealing with highly contrasted scatterers (although the local fluctuations of the acoustical characteristics of the cross-section will be weak in such cases), and present a high-order Born tomographic method named *distorted Born diffraction tomography* [11, 12]. Lastly, the performances and limitations of both tomographic approaches are compared in the basis of experimental data.

The theoretical development of the first-order Born tomographic method is asymptotic, as described using the Green's function in the case of a homogeneous medium. The unknown object function, which has been extended to the case of weakly contrasted and weakly heterogeneous media, is linearly related to the measured field via a Fourier transform, and the inverse problem is related to the filtered back-projections algorithm [13]. In the case of more highly contrasted media, this strategy was adapted experimentally to take into account physical phenomena such as the wave refraction, where the problem can be reduced to the study of a fluid-like cavity buried in an elastic cylinder surrounded by water. Since this iterative experimental method, which is known as *compound ultrasonic tomography*, gives excellent results, the only limitations here are the heavy data processing required and the complex acoustical signals resulting from the multiple physical effects involved.

Nonlinear inversion methods with higher-order levels of approximation have therefore been investigated, including the distorted Born iterative method [14, 15], which is generally applied to solving electromagnetic [16, 17, 18] and optical [19] inverse scattering problems. However, very few ultrasonic experimental reconstructions are available in the literature [20]. The ultrasonic distorted Born diffraction tomography approach based on this iterative method makes it possible to obtain quantitative images. Our theoretical development is algebraic and the iterations are performed numerically by solving the forward and inverse problems at every iteration after calculating an appropriate Green's function; the previous iteration serves in each case to define the surrounding medium with a variable background. Contrary to what occurs with the original approach, distorted tomography requires only a single series of (experimental or numerical) data.

## II. CONFIGURATION AND STATEMENT OF THE PROBLEM

### 1. *Physical considerations*

Since the acoustic impedance of elastic tubes is more highly contrasted than that of the surrounding water (the scatterer is immersed in a water tank), the ultrasonic propagation is perturbed here by the refraction, attenuation and scattering of the waves [21]. The cross-sections of the tube were taken to be isotropic. This scatterer supports the propagation of more complex waves, such as those occurring in elastic volumes, but in this study, it will be modeled like "fluid", since only compressional P-waves and P-P diffraction scattering are taken into account.

Since the frequency range used in the cross-section imaging is [0.15 – 1] MHz, the wavelengths occurring in the tube - typically 2 to 20 mm at compression wave velocities of between 2000 and 3000 m/s - are much larger than the macroscopic porosity and the micro-structural scale of the shell ($\approx$ nm). Our tubes will therefore be assumed to be weakly heterogeneous (i.e., to consist of a homogenized equivalent medium) and the ultrasonic waves in the shell will be less strongly disturbed. The latter assumption is necessary to be able to introduce a linearized propagating theory and to use an asymptotic approximation. On the other hand, the wavelength ([1.5, 10] mm) measured in water ($c_o \approx 1500 m/s$) was similar to the diameter ($\approx$ 10 mm) of the object and the vibration mode was resonance.

## 2. Geometrical considerations

The canonical 2-D geometry we have adopted here is depicted in Figure 1. Let us take an infinite space $\Omega_o$ in a linear homogeneous uncompressible fluid (a water-like substance, for example), named *the background*. Let us also take an elastic cylindrical cavity with an arbitrary cross-section $\Omega_1$ with generators parallel to the z-axis, immersed in the surrounding medium $\Omega_o$ and illuminated by a point source at the point E. We define the position vector $\vec{r}$ in the xy-plane as follows: $\rho_o(\vec{r})$ denotes the density of the surrounding fluid-like medium and the hollow area, $c_o(\vec{r})$ denotes the velocity of the propagating wave, $\rho_1(\vec{r})$ denotes the density of the cavity and $c_1(\vec{r})$ denotes the velocity of the propagating longitudinal wave.

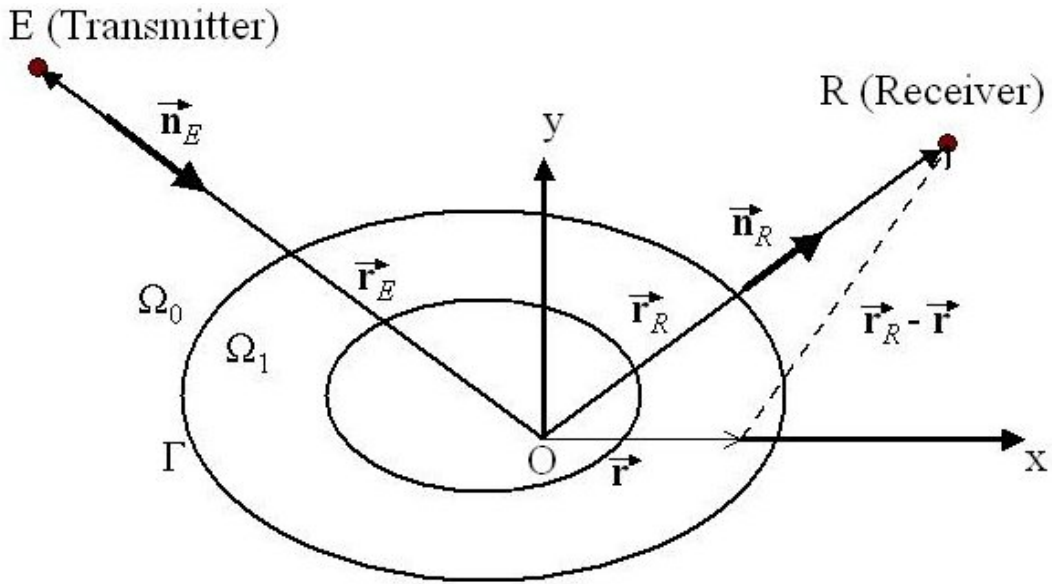

Figure 1: Non-circular cylindrical elastic cavity scattering: geometrical and acoustical configuration

The spatial distribution of the mass density $\rho_o(\vec{r})$, and the velocity $c_o(\vec{r})$ is assumed to be constant everywhere in $\Omega_o$. It will be assumed that an *a priori* value is used for $\rho_1(\vec{r})$.

The aim of this study is to determine the radii and $c_1(\vec{r})$. However, we assume that the radii are restricted to a limited set of values.

### III. ULTRASONIC DIFFRACTED TOMOGRAPHY

The aim of ultrasonic diffraction tomography is to reconstruct the spatial distribution of some geometrical and physical parameters of an object from scattered ultrasonic measurements. These measurements are carried out on variably densely spaced sets of transmitter and receiver positions and the frequencies of an interrogating wave. We are then faced with a forward scattering problem, i.e., predicting the pressure field when the scattering medium and the incident field are assumed to be known, as well as with the inverse scattering problem, i.e., retrieving the parameters of the medium from the measured or simulated incident and scattered fields. This inverse problem is non-linear and ill posed [22, 23], and there is generally no single solution. It is necessary to find a way of eliminating solutions, which do not correspond to reality.

Basic ultrasonic diffraction tomography principles have by now been clearly established in the case of weakly varying media such as low-contrast structures, i.e. almost homogeneous media [24].A constant reference medium can therefore be chosen (i.e., approximations will be made with a constant background). The scattering problem can be linearized by using the first-order Born approximation, and if the Green's function of the unperturbed problem (the background $\Omega_o$) is known, the forward problem can be solved with the Lippmann-Schwinger integral equation and the far field solution of the equation can be calculated. The inverse problem is then solved by performing an asymptotic development and the "classical" tomographic algorithm will yield the perturbation with respect to the reference problem. This leads to a linear relation between the object function (or contrast function) and the scattered field, particularly in the far-field (2-D or 3-D Fourier transform), which makes it possible in principle to reconstruct the object function in almost real time based on a sufficiently large set of scattering data.

If the contrast between the media increases, however, the first-order Born approximation will no longer be valid and we will have to resort to other strategies. The first strategy adopted in this case [25, 9] consists in iteratively correcting the measured or simulated data, depending on the reflection and refraction behavior of the waves propagating through the water/object interface. This strategy, which is known as compound tomography, makes it possible to use the Born approximation, by, "virtually" modifying the Green's function of the scatterer at each iteration, i.e. in each experiment. The disadvantage of this procedure is that as many experiments as iterative steps have to be carried out.

Our second strategy involves the algebraic inversion of the scattered field, based on the distorted Born iterative method, using iterative numerical steps and just one experiment.

### A. First-order Born tomography

In the case of our configuration, we take $\Omega_o$ to denote the known part (the background) identified by the density $\rho_0$ and the velocity $c_0$, and $\Omega_1$ to denote the unknown part (the perturbation) identified by the (variable) density $\rho_1(\vec{r})$ and the (variable) velocity $c_1(\vec{r})$.

In the case of weak scattering ($\rho_1(\vec{r}) c_1(\vec{r}) \approx \rho_0 c_0$), the temporal equation that best describes the acoustic propagation/diffusion processes occurring in the medium (including the boundary and Sommerfeld conditions) results from the Pekeris equation and is given by:

$$\text{Equation 1:} \quad k_0^2 p_a + \Delta p_a = (k_0^2 - k_1^2) p_a - S_p + \overrightarrow{\text{grad}}\left(\log\left(\frac{\rho_1}{\rho_0}\right)\right) \cdot \overrightarrow{\text{grad}}\, p_a$$

where $p_a$ is the acoustic pressure, $k_0$ and $k_1(\vec{r})$ are the complex wave numbers.

Upon introducing $f(\vec{r}) = k_1^2(\vec{r}) - k_0^2 - \eta^{-1}(\vec{r})\Delta\eta(\vec{r})$, the contrast function, $p = \eta p_a$, with $\eta = \left(\frac{\rho_1}{\rho_0}\right)^{-1/2}$ and $q = \eta S_p$, we obtain:

$$\text{Equation 2:} \quad \Delta p + k_0^2 p = -f(\vec{r})\, p - q$$

If the Green's function of the non-perturbed problem is known, we can determine the total pressure field in $\Omega_o$ using the Lippmann-Schwinger integral equation:

$$\text{Equation 3:} \quad p(\vec{r}) = p_i(\vec{r}) + \int_{RO} G_0(\vec{r}, \vec{r}')f(\vec{r}')p(\vec{r}')d\vec{r}'$$

To determine the pressure in the heterogeneous region (RO), the Lippmann-Schwinger equation must be solve but the problem is a non-linear. The intuitive solution consists in linearizing the problem using the Born approximation. We assume that the diffracted field is negligible in comparison with the incident field in RO, which is valid in the case of weakly heterogeneous media:

$$\text{Equation 4: } f(\vec{r}) = k_1^2(\vec{r}) - k_0^2 - \eta^{-1}(\vec{r})\Delta\eta(\vec{r}) \cong 0$$

Upon introducing $f^0 = f$ under the Born approximation, we obtain the following forward problem:

$$\text{Equation 5: } p_s(\vec{r}) = p(\vec{r}) - p_i(\vec{r}) = \int_{RO} G_0(\vec{r}, \vec{r}\,') f^0(\vec{r}\,') p_i(\vec{r}\,') d\vec{r}\,'$$

Let us assume that the incident field is a plane wave (in the far field of the transducer):

$$\text{Equation 6: } p_i(\vec{r}_E, \vec{r}) = A \exp(ik_0 \vec{n}_E \cdot \vec{r})$$

where $\vec{n}_E$ is the incident normal vector.

Within the framework of the Born approximation, the scattered field measured by a receiver at point $\vec{r}_R$ is:

$$\text{Equation 7: } p_s(\vec{r}_R) = \int_{RO} G_0(\vec{r}_R, \vec{r}) f^0(\vec{r}) p_i(\vec{r}) d\vec{r}$$

with

$$\text{Equation 8: } G_0(\vec{r}_R, \vec{r}) \approx \frac{i}{4}\sqrt{\frac{2}{\pi k_0 |\vec{r}_R|}} \exp\left(ik_0 |\vec{r}_R| - \frac{\pi}{4}\right) \exp\left(-ik_0 \vec{n}_R \cdot \vec{r}\right)$$

where $\vec{n}_R$ is the observation normal vector.

The far field Born approximation of the scattered field is given by:

Equation 9:

$$p_s(\vec{r}_R) = \int_{RO} \frac{i}{4}\sqrt{\frac{2}{\pi k_0 |\vec{r}_R|}} \exp\left(ik_0 |\vec{r}_R| - \frac{\pi}{4}\right) \exp\left(-ik_0 \vec{n}_R \cdot \vec{r}\right) f^0(\vec{r}) A \exp\left(ik_0 \vec{n}_E \cdot \vec{r}\right) d\vec{r}$$

Equation 10: $p_s(\vec{r}_R) = A \dfrac{i}{4}\sqrt{\dfrac{2}{\pi k_0 |\vec{r}_R|}} \exp\left(ik_0|\vec{r}_R| - \dfrac{\pi}{4}\right) \int_{RO} f^0(\vec{r}) \exp\left(-ik_0(\vec{n}_R - \vec{n}_E).\vec{r}\right) d\vec{r}$

If we perform measurements on a circle surrounding the object, we obtain:

Equation 11: $p_s(\vec{r}_R) = B \int_{RO} f^0(\vec{r}) \exp\left(-ik_0(\vec{n}_R - \vec{n}_E).\vec{r}\right) d\vec{r}$

where B is known and depends only the frequency when the radius of the circle is given. This equation is the 2-D Fourier transform of the contrast function $f^0$. The inverse problem, i.e. the restitution of $f^0$, is an inverse Fourier transform. The spectral aperture of the frequency space depends on $k_0(\vec{n}_R - \vec{n}_E)$ and is therefore not complete. This aperture is then experimentally limited at several points inside a limited bandwidth/range/field depending on the frequency range of the transducers. Furthermore, the coverage of the Fourier plane of the object by the data obtained is only partial and it is therefore necessary to introduce interpolations between the projections to be able to use a classical 2-D Fourier transform. Our algorithms are based on a special angular and linear scanning process performed on the radial measurements of the spatial Fourier transform [26]. It therefore suffices to cross the Cartesian coordinates with the polar coordinates in the Fourier space to be able to carry out the reconstruction without having to interpolate the object. This is done using the algorithm of reconstruction by summation of filtered back-projections [27, 28].

### B. Limits in the case of a high-impedance contrast scatterer

In the ultrasonic characterization of high-impedance contrast scatterers, it is not possible to use the first-order Born approximation with a constant background. Bodies of this kind are quite heterogeneous, and their acoustical characteristics are quite different from those of the surrounding soft medium.

If we assimilate the present tube imaging problem to the identification of a water-like cavity present in an elastic cylinder, and adopt the previous acoustical assumptions about the weak heterogeneity of the shell and the propagation, the first-order Born approximation with a variable background can be used if the background is defined as the fixed, solid part without any hollow (the water) and the perturbation, i.e. the object to be reconstructed, is defined as the cavity. But there will be a bias in the thickness measurements and shape imaging. In fact, "classical" reconstruction tools based on the first-order Born approximation assume the occurrence of straight ray propagation, whereas a large deviation of the incident beam occurs at the interface between the water and the tube, due to the difference between the acoustic properties of the two media. Due to the refraction effects, diffraction occurs on the irregularly shaped scatterer. Our approach was designed first to cancel out the refraction effects by using a specific experimental set-up in order to impose straight ray propagation inside the shell during high-frequency (> 500 kHz) tomographic measurements [9, 29]. This procedure requires exactly determining the position and the geometrical shape of the scatterer. Our initial attempts on these lines have been improved using signal and image processing methods [30], but this strategy is still too complicated to be suitable for experimental purposes.

The solution proposed in this paper consists in iteratively calculating the Green's function of the inhomogeneous problem beginning with the first-order Born approximation previously defined. The first perturbation operator is the same as with a homogeneous background and all the perturbation parameters are defined in terms of the characteristics of the medium. The first results obtained were similar to those obtained in the case of a constant background, with various effects at the water/shell interface. In the next step, the inverse problem is solved using a procedure to solve the forward problem whereby the contrast function *f* of the scatterer has been determined at the previous step.

### C. Distorted Born diffraction tomography

The distorted Born diffraction tomography is based on the distorted Born iterative (DBI) method. This approach involves performing iterative minimization on the discrepancy between actual measurements and the integral representation of the field obtained by solving a forward problem with an estimated contrast function.

The integral expression is discretized using the moment method. The inhomogeneous region (RO) is replaced by an array of elementary square cells $\Omega_{pq}$ (Figure 2) which are small enough for $k(\vec{r})$ and $p(\vec{r})$ to be presumably constant therein. This leads to a discretized form of Equation 3:

$$\text{Equation 12:} \quad p(\vec{r}) = p_i(\vec{r}) + \sum_{p=1}^{n}\sum_{q=1}^{n} f(\vec{X}_{pq})\, p(\vec{X}_{pq}) \int_{\Omega_{pq}} G(\vec{r}, \vec{r}')\, d\vec{r}'$$

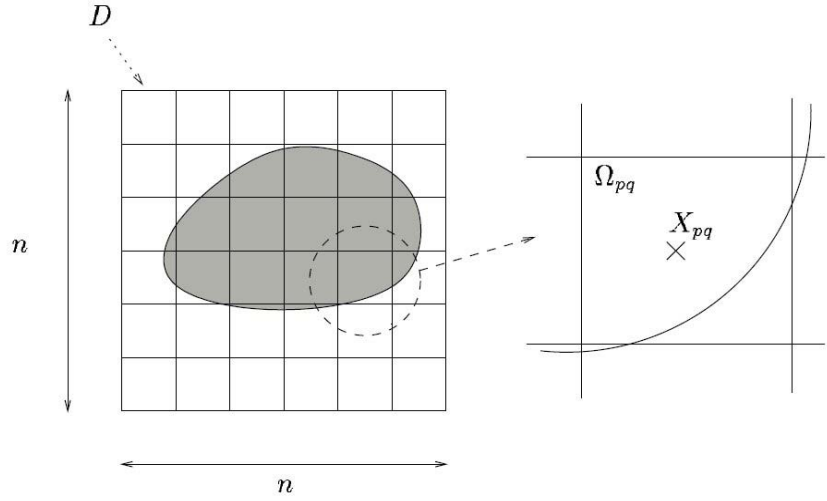

Figure 2: Discretization of the object

In the *forward problem,* the contrast function and the incident wave are known. The first step consists in finding the pressure field $p(\vec{X}_{pq})$ in (RO). The scattered pressure field can then be calculated elsewhere via Equation 12. $p(\vec{X}_{pq})$ is determined using the collocation method. Writing Equation 12 for all points $\vec{r} = \vec{X}_{\alpha\beta}$ $\alpha, \beta = 1,2,...,n$, we obtain a system of $n^2$ linear equations in terms of $n^2$ unknowns:

Equation 13: $$p_i(\vec{X}_{\alpha\beta}) = \sum_{p=1}^{n} \sum_{q=1}^{n} \left[ \delta_{\alpha p} \delta_{\beta q} - f(\vec{X}_{pq}) \int_{\Omega_{pq}} G(\vec{X}_{\alpha\beta}, \vec{r}') d\vec{r}' \right] p(\vec{X}_{pq})$$

This can also can be expressed as the linear matrix equation: $P_i = \Gamma(F) P$ where

$P_i = \begin{bmatrix} p_i(\vec{X}_{11}) \\ . \\ p_i(\vec{X}_{nn}) \end{bmatrix}$, $P = \begin{bmatrix} p(\vec{X}_{11}) \\ . \\ p(\vec{X}_{nn}) \end{bmatrix}$ the incident and total pressure fields and $F = \begin{bmatrix} f(\vec{X}_{11}) \\ . \\ f(\vec{X}_{nn}) \end{bmatrix}$ the contrast function.

In the *inverse problem*, the incident field is known, the total field is known at some $M$ measurements points $\vec{r}_j$, $j = 1,2,...,M$; and we have to determine the unknown contrast function $F$ in (RO). Via Equation 12, the inverse problem can be expressed as a non-linear system of $M$ equations in terms of the $n^2$ unknowns $f(\vec{X}_{pq})$ :

Equation 14: $p_s(\vec{r}_j) = \sum_{p=1}^{n} \sum_{q=1}^{n} f(\vec{X}_{pq}) p(\vec{X}_{pq}) \int_{\Omega_{pq}} G(\vec{r}_j, \vec{r}') d\vec{r}'$, $j = 1...M$

This can also can be written in the form: $P_s = Q(P) F$, with $P_s = \begin{bmatrix} p_s(\vec{r}_1) \\ . \\ p_s(\vec{r}_M) \end{bmatrix}$

To solve this equation, we use the distorted Born iterative method (DBI), which consists in solving the problem using successive linear estimates and updating the total field and the Green's function at each iteration. The initial guess is provided by the first-order Born solution $F^0$. Then, if the *l*-order solution $f^l$ is assumed to be known, the $(l+1)$-order solution verifies [12, 33]:

Equation 15: $p_s(\vec{r}_j) - p_s^l(\vec{r}_j) = \sum_{p=1}^{n} \sum_{q=1}^{n} \left\{ \left( f^{l+1}(\vec{X}_{pq}) - f^l(\vec{X}_{pq}) \right) p^l(\vec{X}_{pq}) \int_{\Omega_{pq}} G^l(\vec{r}_j, \vec{r}') d\vec{r}' \right\}$

Where $G^l$ is the inhomogeneous Green's function of the object $f^l$ immersed in $\Omega_0$. $p^l$ and $p_s^l$ are the total and scattered pressure fields for this scatterer $f^l$ illuminated by a source point placed at E (Figure 1). These quantities are calculated using the integral representation (12) and by solving a forward problem for the known scatterer $f^l$. The DBI procedure can be written in condensed form as follows:

$$\text{Equation 16:} \begin{cases} P^0 = P_i \\ P_s = Q(P_i) F^0 \\ P^l = [\Gamma(F^l)]^{-1} P_i \\ P_s - P_s^l = Q^l(P^l)[F^{l+1} - F^l] \end{cases}$$

The iterative procedure is continued until $\dfrac{\|F^{l+1} - F^l\|}{\|F^l\|}$ is smaller than $\varepsilon$ (which is taken to be 5%).

It is worth noting that $Q^l$ is generally a non-square and ill-conditioned matrix. A mean-square solution of the inverse problem is obtained by performing a Singular Values Decomposition (SVD) associated with a Tikhonov regularization procedure [31].

## IV. WATER TANK MEASUREMENTS

The experimental setup used here was designed for performing diffraction measurements. The acoustic device involved is able to move with various degrees of freedom and used to analyze samples in all directions. The positions of the target and the transducers can be adjusted. In particular, the operator can impose precise rotations and translations on the transmitter and the receiver.

Ultrasounds were generated using a pulse receiver device and piezo-composite wide-band transducers. The nominal frequency of the transducers was 0.25 MHz and the usable bandwidth of the transducers was approximately 135 to 375 kHz.

The object was placed in the center of the measuring system. Ultrasonic measurements were performed in water at room temperature. The sector scanned was 360° with both transducers, and the angular increment was "5-degree" (72 x 72 signals). Transmitted and received ultrasound Radio-Frequency (RF) signals were digitized (8 bits, 20 MHz) using an oscilloscope, stored (4000 samples) on a computer via a General Purpose Interface Bus for off-line analysis.

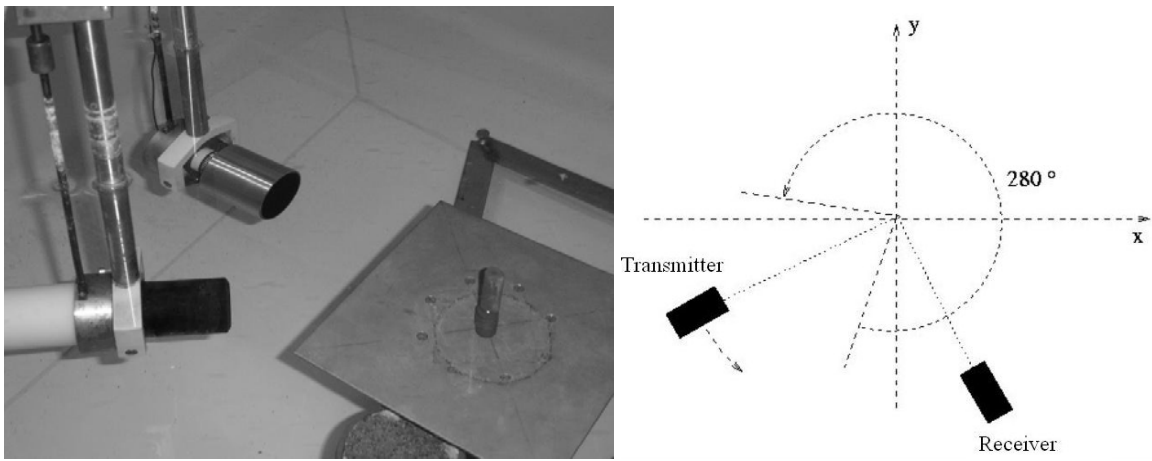

Figure 3: Water tank and experimental configuration

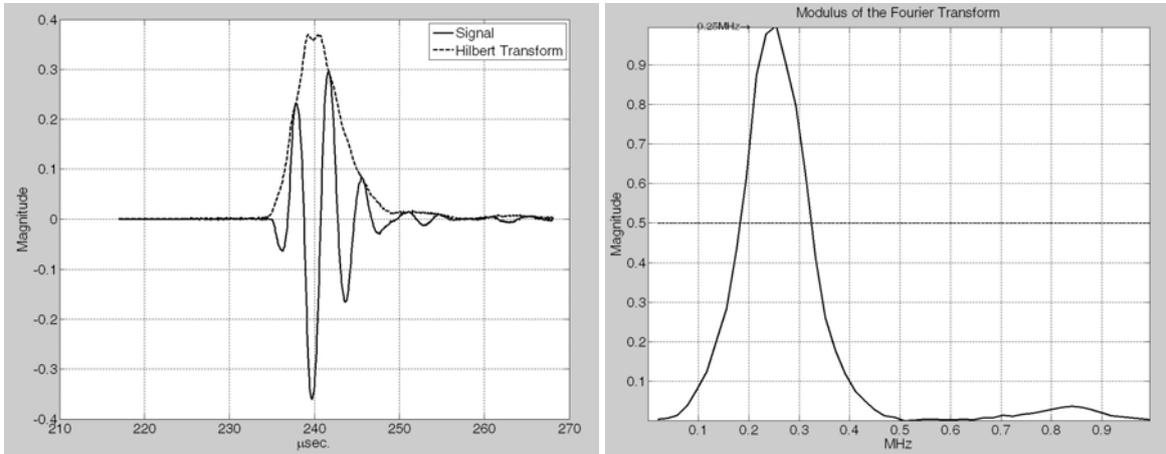

Figure 4: Experimental pulse response (left) and transfer function (right) of the transmitter

(Nominal frequency: 250 kHz, Sample frequency: 20 MHz)

A circular data recording in the reflected mode (Figure 5) on a very thin copper wire (diameter 0.07 mm) was used to estimate the (mechanical deviation of) difference in the position and the time of flight between the transducers and the center of the bench.

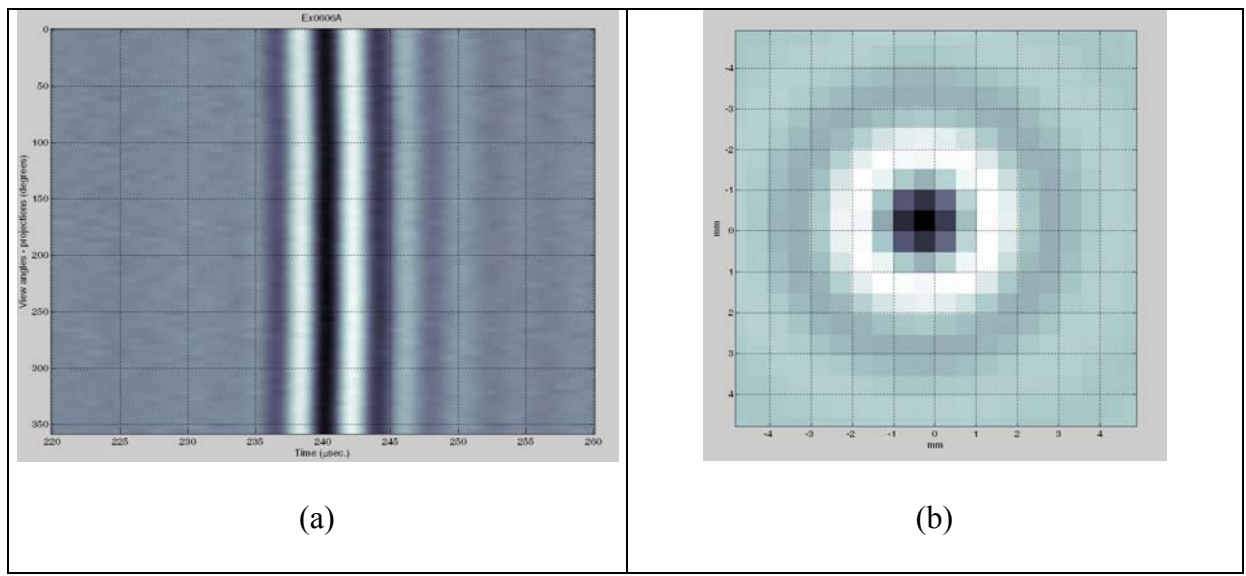

(a)   (b)

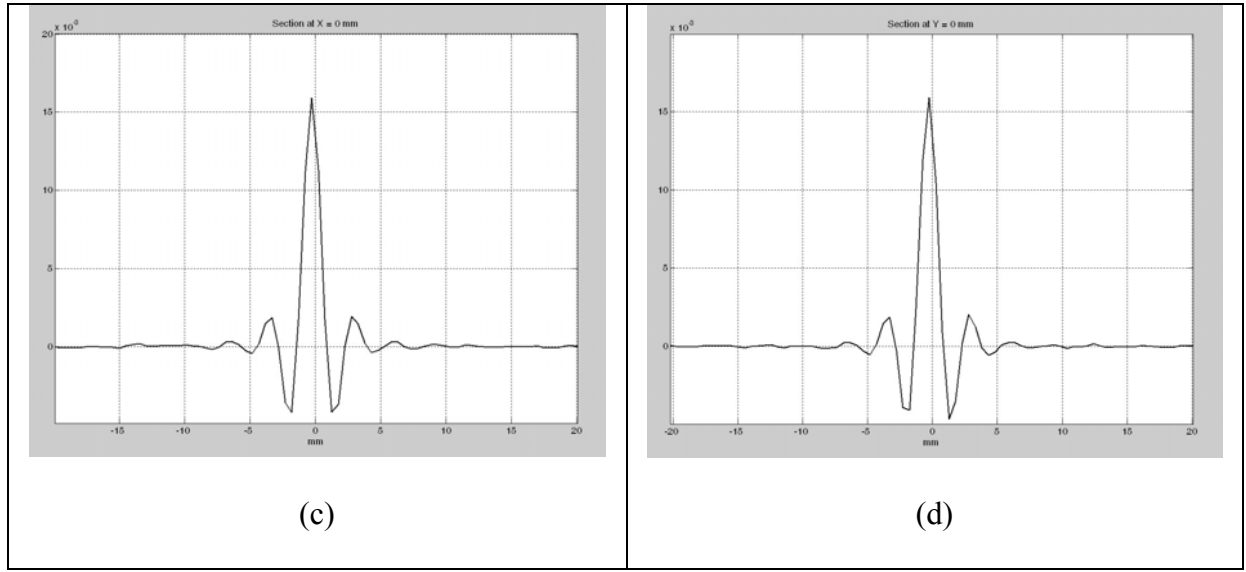

(c)  (d)

Figure 5: Ultrasonic first-order *Born* reflection tomography of a thin copper wire (0.07 mm), (a) sinogram, 72 back-projections through 360°, 1024 samples (b) tomogram 300 x 300 pixels – zoom (c, d) x and y pixel profiles drawn at x = 0 mm and y = 0 mm

## V.    NON-CIRCULAR ELASTIC TUBE

The experimental sample used here (Figure 6) was a non-circular homogeneous isotropic shell made of artificial resin (NEUKADUR ProtoCast 113™). Its maximum internal and external diameters were measured with a caliper and were respectively 6 and 12 mm. The density of the resin was $\rho_1 \approx 1150 \ kg/m^3$, and the mean velocity of the compressional wave was $c_1 \approx 2400 \pm 50$ m/s. The surrounding fluid-like medium (and the hollow area) was water at a temperature of 18 ° ($\rho_o$ = 1000 $kg/m^3$, $c_o$ = 1480 m/s). The transmitter and the receiver were placed 17.5 cm to the right of the center of the bench.

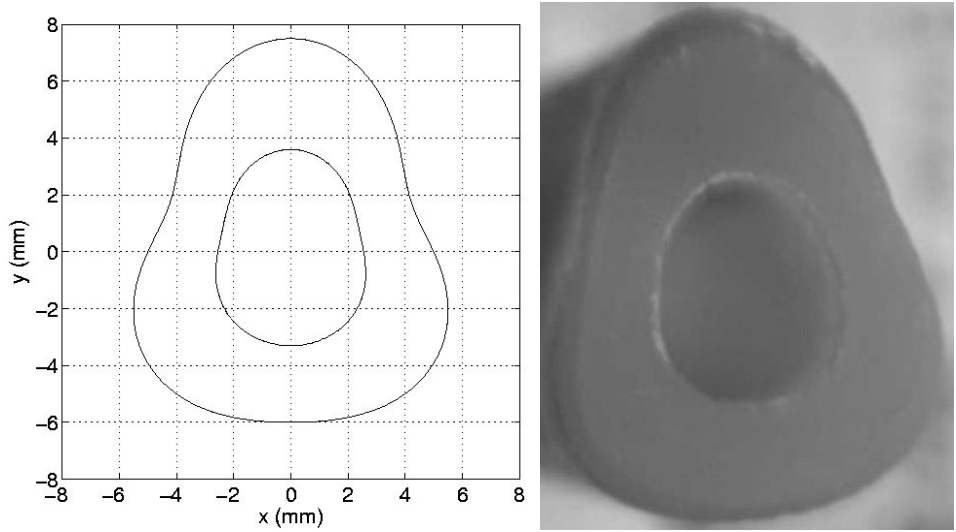

Figure 6: NEUKADUR ProtoCast 113[TM] resin tube dimensions (left),

sample (right)

### A. Simulations

We first tested the efficiency of the distorted Born iterative algorithm on simulated values generated using a boundary integral equation (BIE) method [32]. The transmitter was rotated in "5-degree" steps and the receiver, in 10- degree steps. Five frequencies were used for the simulations: 150 kHz, 180 kHz, 250kHz, 300 kHz and 350 kHz. Results are shown in Figure 7. In this simulation, we assumed the densities $\rho_1$ and $\rho_o$ to be identical. The contrast function $f$ therefore depended only on the compressional wave velocity.

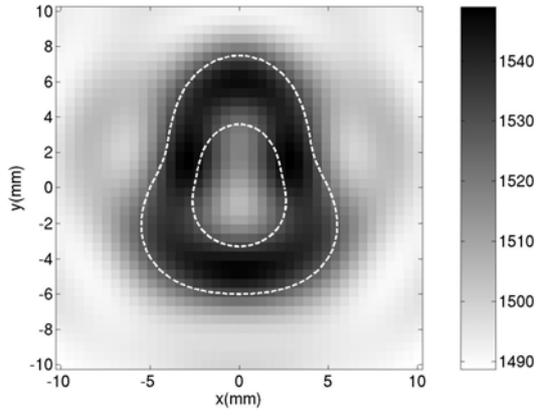
(a)

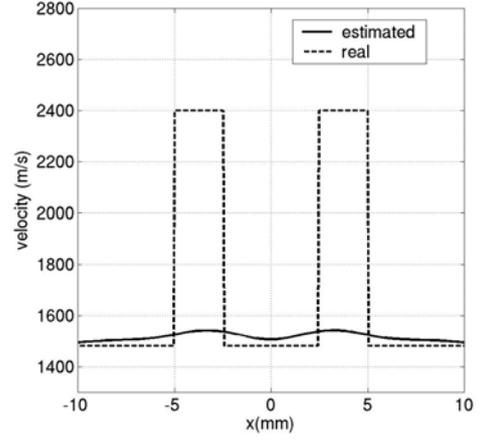
(b)

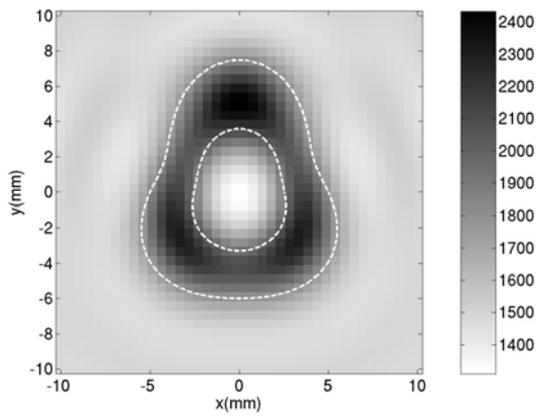
(c)

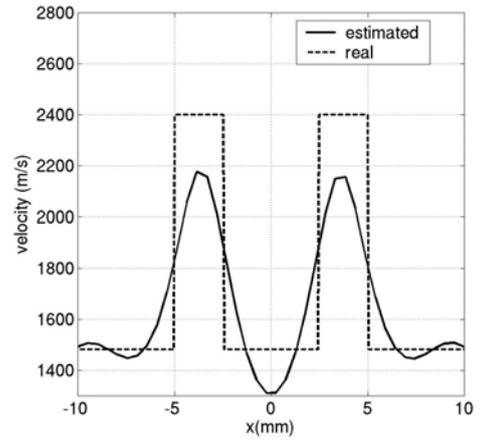
(d)

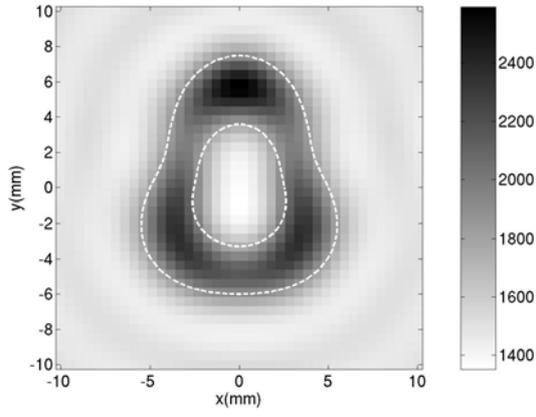

(e)

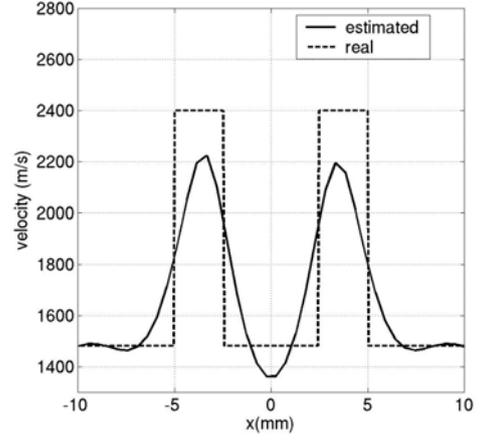

(f)

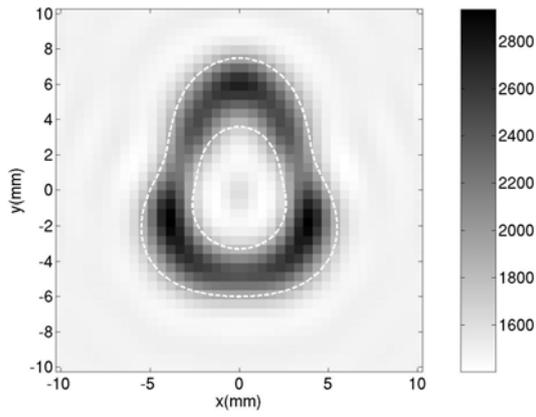

(g)

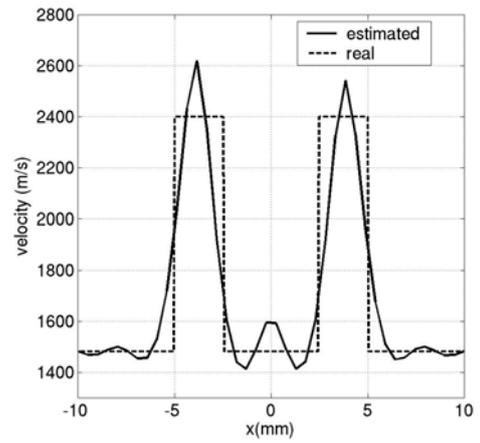

(h)

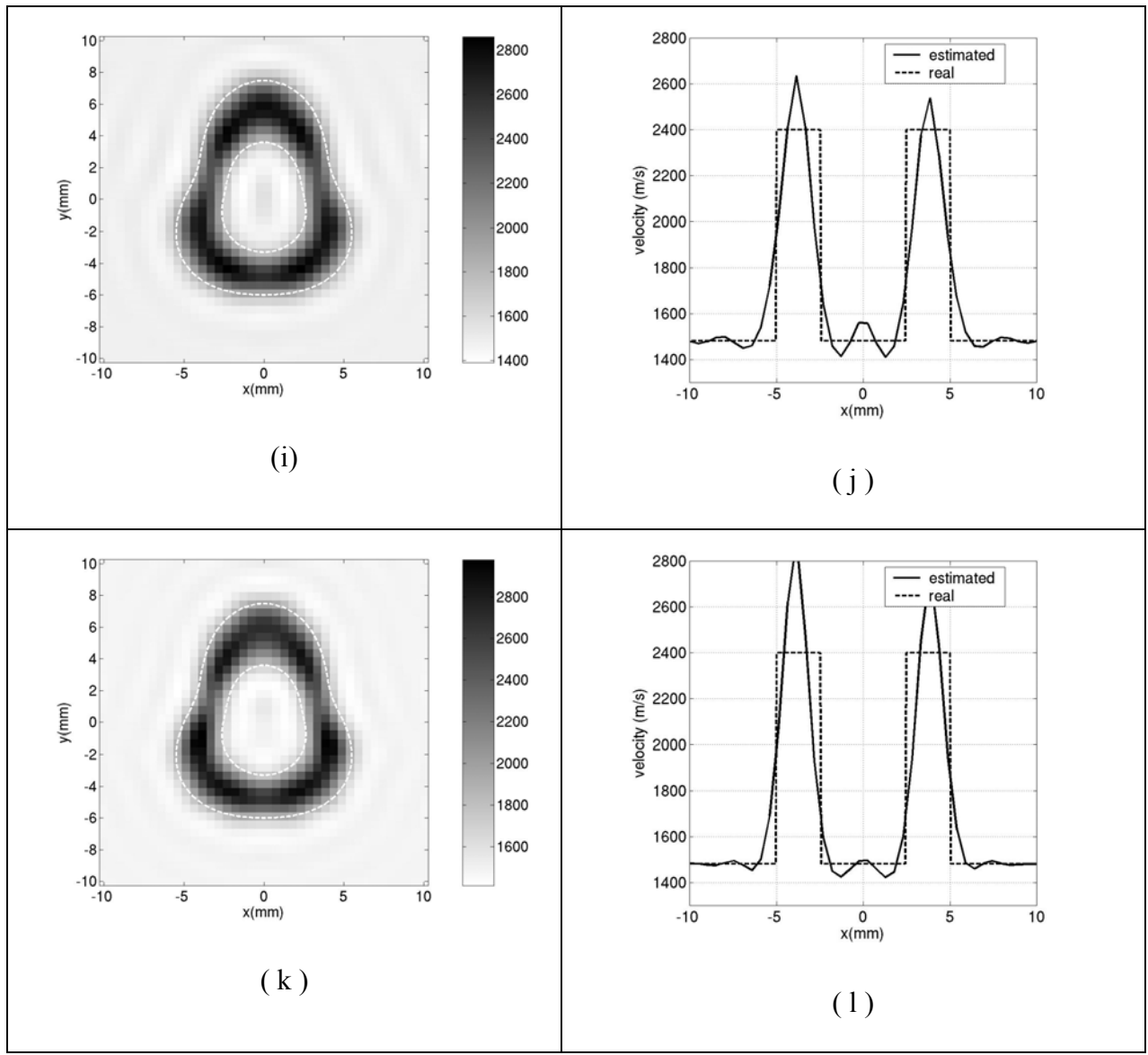

Figure 7: Ultrasonic distorted Born iterative tomography of the non-circular elastic tube. Simulated data, left: tomograms 40 x 40 pixels, right: x-pixel profiles drawn at y = 0 mm; (a – b) initial solution at 150 kHz; (c - d) 17 iterations at 150 kHz; (e – f) 11 iterations at 180 kHz; (g - h) 19 iterations at 250 kHz; (i – j) 13 iterations at 300 kHz; (k – l) 12 iterations at 350 kHz

The initial frequency was chosen so that the first-order Born solution does not include any artifacts. Previous studies [33] have shown that if the phase shift resulting from the presence of the scatterer is greater than π, the model will include large artifacts. We therefore chose a frequency such as $f \langle \frac{c_0 c_1}{2d|c_1 - c_0|} \approx 150 \,\text{kHz}$ (where d is the largest dimension of the scatterer). The initial solution is given by the first-order Born estimate. At each frequency, the initial solution was the final result of the iterations performed at the previous frequency. The numerical simulations were found to be satisfactory on the qualitative level (as regards the shape, dimensions, and location ). From the quantitative point of view (velocity measurements), the algorithm was found to overestimate the velocity, which depends strongly on the regularization parameter used for the mean-square inversion procedure. A study is now in progress with a view to optimizing this parameter [34, 35].

### B. Experimental

We then compared the performances of the first-order Born tomography and the distorted Born diffraction tomography with a set of experimental data. With the transducers used, which had a nominal frequency of 250 kHz, the frequency range was limited to the [135, 375] kHz bandwidth. The image resolution was therefore limited by frequency range available.
Figure 8 shows the first-order Born *reflection* tomography of the scatterer. No corrections or signal processing have been carried out on these images. The sinogram corresponds to the modulus of the Hilbert transform of the initial RF-signals. Only one transducer was used, and it was placed in 72 positions around the object with a 5- degree step. Figure 9 shows the first-order Born *diffraction* tomography of the scatterer. Here again, no corrections or signal processing have been carried out on these images. The sinogram corresponds to the modulus of the Hilbert transform of the initial RF-signals. The two transducers used were placed in 72 by 72 positions around the object with a 5-degree step

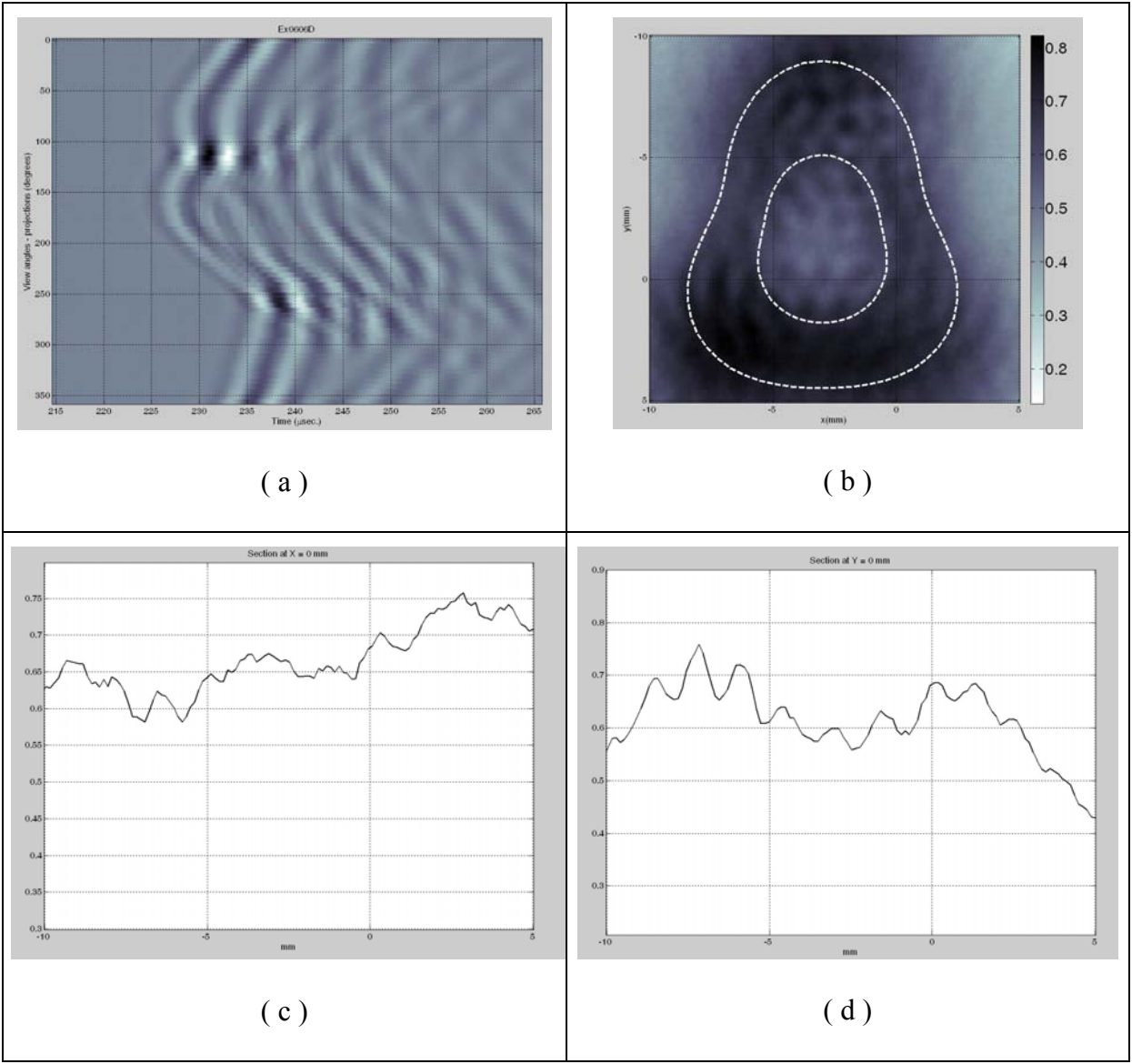

(a)

(b)

(c)

(d)

Figure 8: Ultrasonic first-order Born *reflection* tomography of the non-circular elastic tube. Experimental data. (a) sinogram, Hilbert transform of RF-signals, 72 back-projections

through 360°, 1024 samples; (b) tomogram with 300 x 300 pixels – zoom; (c, d) x and y pixel profiles drawn at x = 0 mm and y = 0 mm

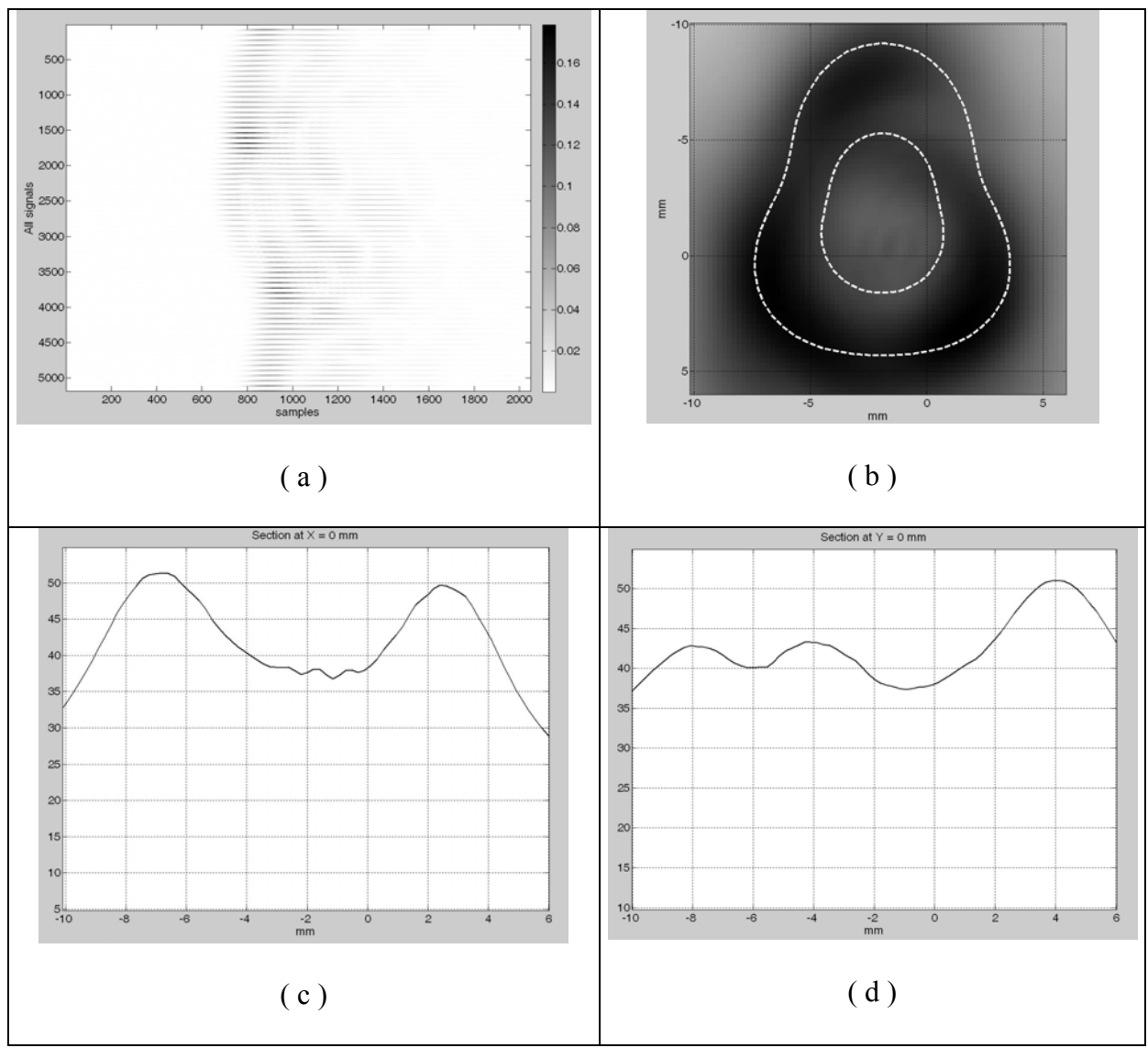

(a)

(b)

(c)

(d)

Figure 9: Ultrasonic first-order Born *diffraction* tomography of the non-circular elastic tube. Experimental data. (a) sinogram, Hilbert transform of RF-signals, 72 x 72 projections through 360°, 1024 samples; (b) tomogram with 300 x 300 pixels; (c, d) x and y pixel profiles drawn at x = 0 mm and y = 0 mm

Figure 10 shows the distorted Born iterative diffraction tomography of the scatterer. No corrections or signal processing were carried out on these images. The experimental configuration is described in section IV. The sinogram correspond to that previously presented (Figure 9-a). The scattered field was obtained by subtracting the incident field (measured without a target) from the total field. The frequency data were obtained using the numerical Fast Fourier Transform of the temporal data.

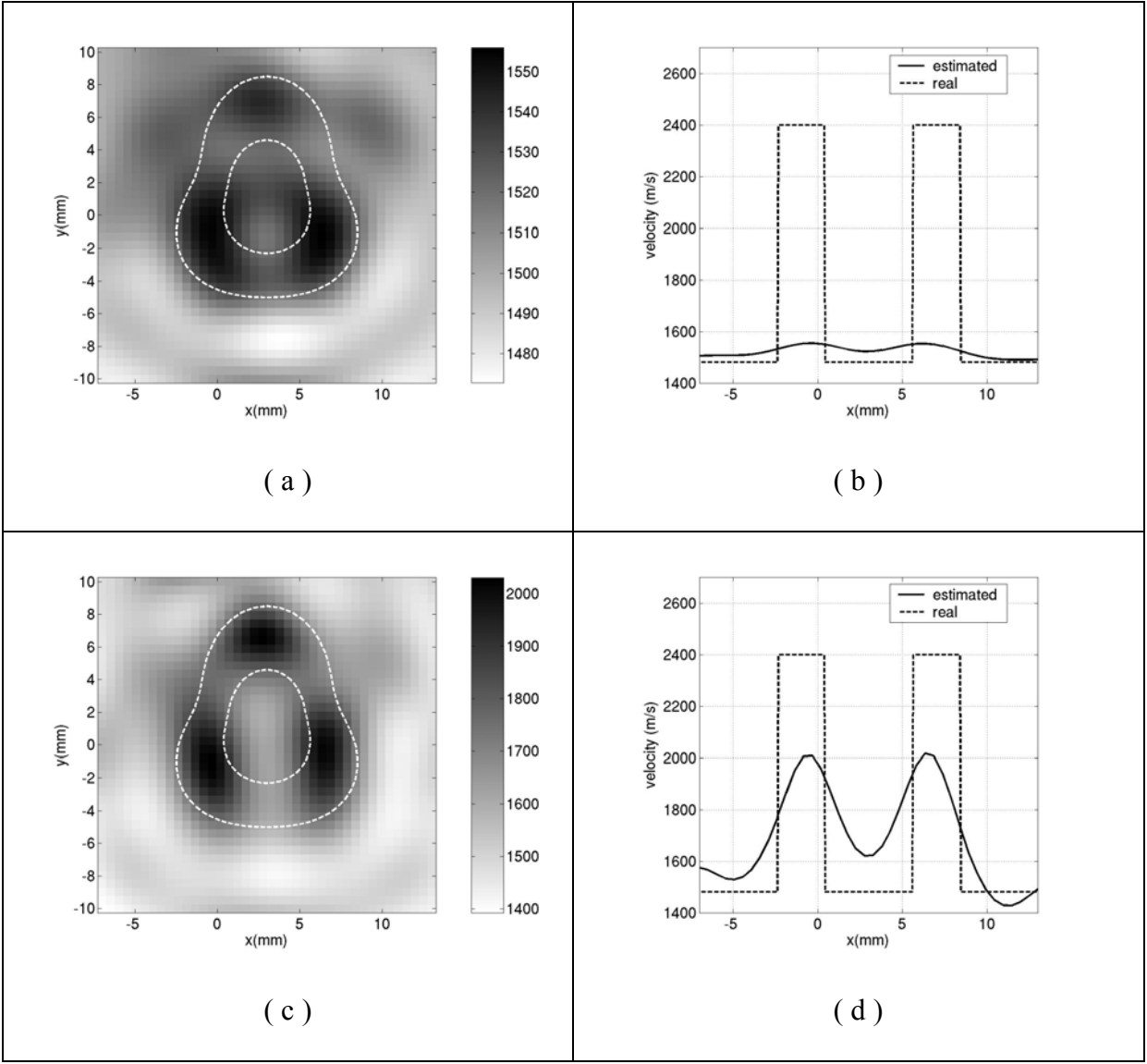

( a )

( b )

( c )

( d )

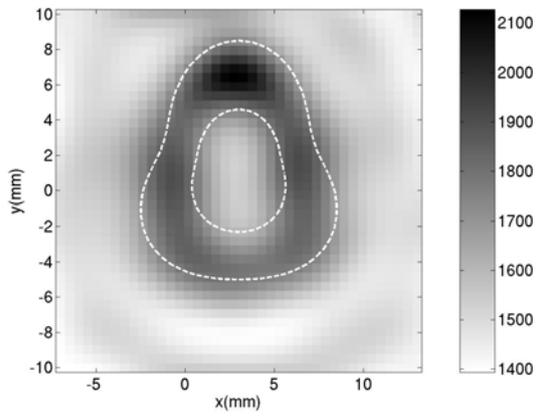

( e )

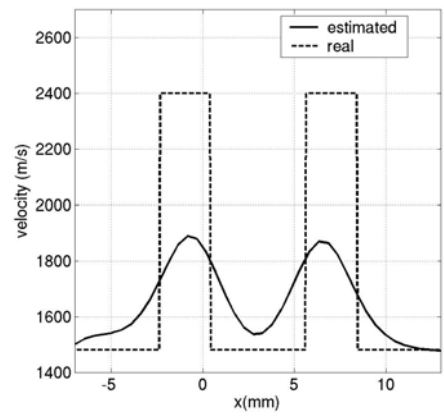

( f )

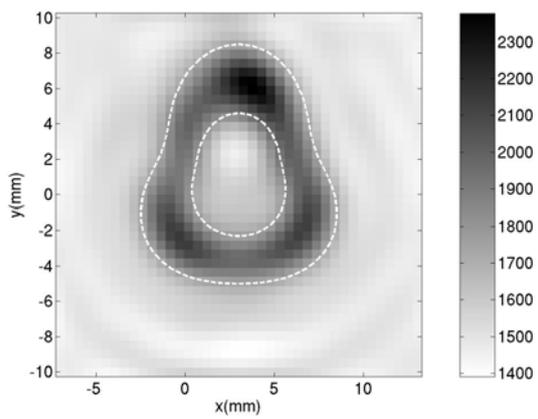

( g )

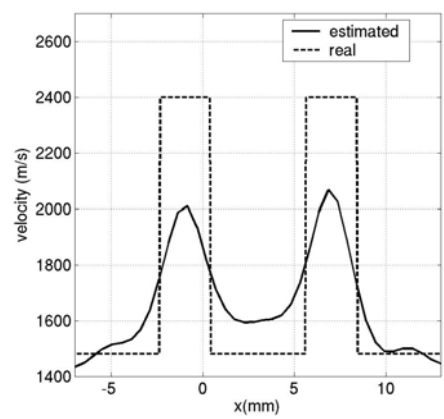

( h )

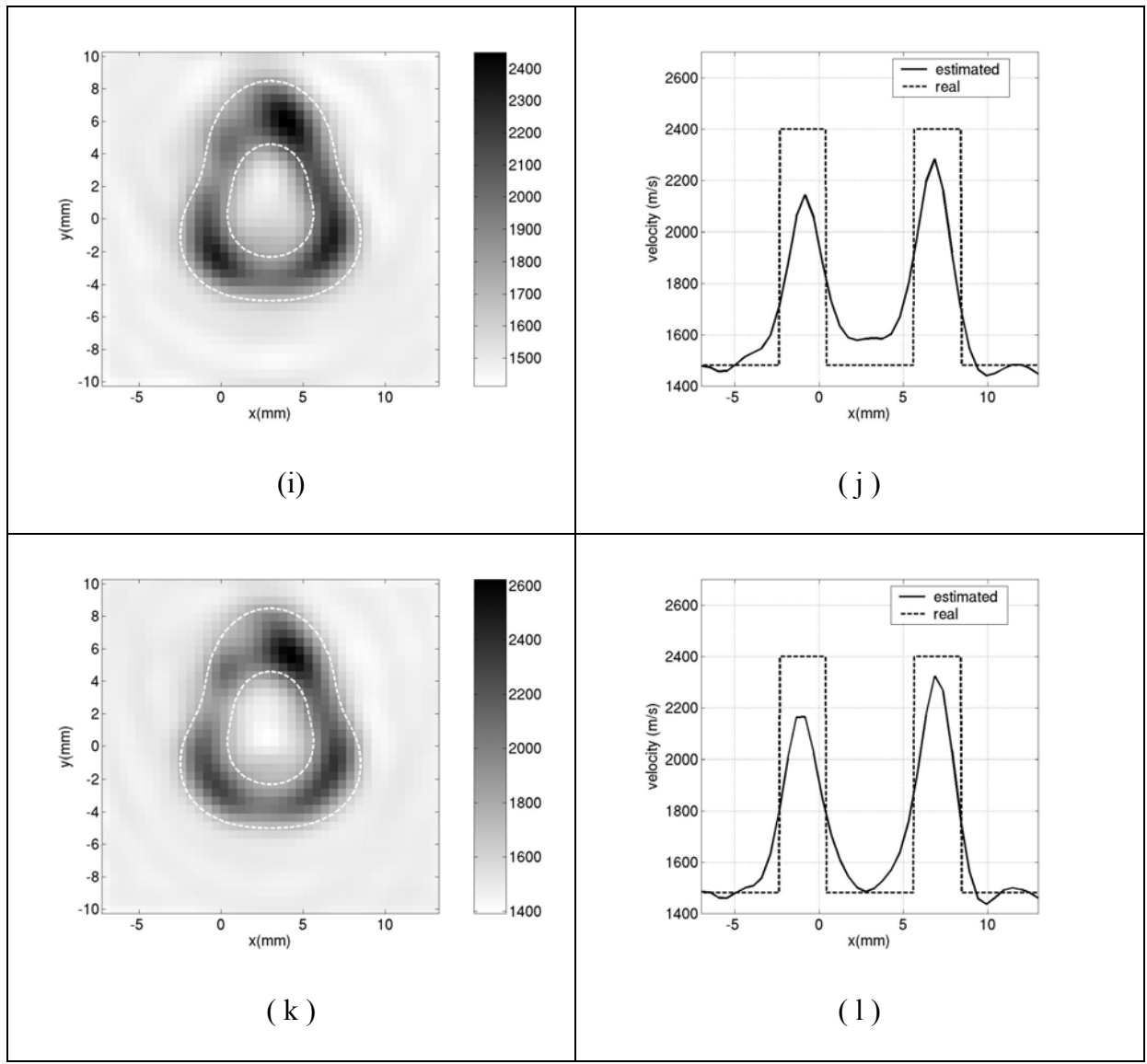

Figure 10: Ultrasonic distorted Born diffraction tomography of the non-circular elastic tube. Experimental data. (a – b) initial solution at 150 kHz; (c - d) 13 iterations at 150 kHz; (e – f) 9 iterations at 180 kHz; (g - h) 17 iterations at 250 kHz; (i – j) 15 iterations at 300 kHz; (k – l) 11 iterations at 350 kHz

Figure 8 and Figure 9 give the results obtained using the classical back-projection algorithm. In this algorithm based on the first-order Born approximation, all the available frequencies between 135 kHz and 375 kHz were used. Because of the large impedance contrast and since the wavelength was similar to the size of the scatterer, the image is bound to include considerable artifacts and the resolution is not satisfactory. In addition, the image obtained here does not give any quantitative information about the speed of sound. The first-order Born tomography was based on a single-scattering theory and there will be a bias in the assessment of the shell thickness. It only gives a qualitative picture of the "perturbation", i.e. of the external boundaries.

Figure 10 gives the results obtained using the distorted Born diffraction tomography. First it is worth noting that in our experimental data, differences in the density were observed between the target and the surrounding medium that are not taken into account in the inversion scheme. It can be seen here that the resolution and the quality of the modeled contrast improved gradually, as with the inversion of the numerical data (Figure 7). The final result of the iteration process was satisfactory, even as far as the value of the contrast was concerned. As in the case of the model, the geometry was fairly accurate, whereas the velocity was less accurately estimated, with a relative error of about 5 %. As in the case of the simulations, this was mainly due to the regularization parameter selected.

## VI. CONCLUSION

This paper deals with the two-dimensional imaging of a non-circular elastic tube using *distorted Born iterative* tomography. The scope of the ultrasonic *Born* tomography was thus extended from low impedance contrast media (the classical domain) to higher impedance contrast domains, using a correction scheme with the following features: the wave propagation and the associated Green's function of the medium are determined at the previous step. The ultrasonic propagation is greatly perturbed by the difference in the acoustic impedance between the cross-section of the scatterer and the surrounding homogeneous reference material, which generates large parasite processes. The strategy used to solve the problem was based on the distorted Born iterative method, where the iterations were performed numerically, based on a single experimental measurement. This strategy did not require any *a priori* information to give a reasonable contrast, but this information was more necessary for the iteration procedure. The examples given in this paper involve a non-canonical homogeneous shape, and the results were based on numerical simulations and experiments. The results obtained here are most promising because the geometrical and physical parameters of the scatterer were accurately reconstructed.

Various way of improving this method will now be investigated. One of the most important aspects of the inversion scheme is the choice of regularization procedure. In this paper, the Tikhonov regularization method was used, and work is in progress on how to choose the regularization parameter. This classic regularization method yields promising good results, but one of its drawbacks is that the whole image is smoothed with this method, and the edges are not accurately reconstructed. To solve this problem and improve the quality of the images (especially with highly contrasted targets), other "edge-preserving" regularizing methods will have to be employed [36, 37]